\documentclass[prd,superscriptaddress,unsortedaddress,twocolumn,showpacs,preprintnumbers,amsmath,amssymb]{revtex4}

\usepackage[dvips]{graphicx}
\usepackage{amsmath,amssymb,times}
\usepackage{ulem}

\def\lessa{\mathrel{\mathpalette\fun <}}
\def\greata{\mathrel{\mathpalette\fun >}}
\def\fun#1#2{\lower3.6pt\vbox{\baselineskip0pt\lineskip.9pt
\ialign{$\mathsurround=0pt#1\hfil##\hfil$\crcr#2\crcr\sim\crcr}}}

\newcommand{\beq}{\begin{equation}}
\newcommand{\eeq}{\end{equation}}
\newcommand{\bea}{\begin{eqnarray}}
\newcommand{\eea}{\end{eqnarray}}

\DeclareSymbolFont{boldletters}{OML}{cmm} {b}{it}
\DeclareSymbolFontAlphabet{\mathbit}{boldletters}
\DeclareMathSymbol{\alpha}{\mathalpha}{letters}{"0B}
\DeclareMathSymbol{\beta}{\mathalpha}{letters}{"0C}
\DeclareMathSymbol{\gamma}{\mathalpha}{letters}{"0D}
\DeclareMathSymbol{\delta}{\mathalpha}{letters}{"0E}
\DeclareMathSymbol{\epsilon}{\mathalpha}{letters}{"0F}
\DeclareMathSymbol{\zeta}{\mathalpha}{letters}{"10}
\DeclareMathSymbol{\eta}{\mathalpha}{letters}{"11}
\DeclareMathSymbol{\theta}{\mathalpha}{letters}{"12}
\DeclareMathSymbol{\iota}{\mathalpha}{letters}{"13}
\DeclareMathSymbol{\kappa}{\mathalpha}{letters}{"14}
\DeclareMathSymbol{\lambda}{\mathalpha}{letters}{"15}
\DeclareMathSymbol{\mu}{\mathalpha}{letters}{"16}
\DeclareMathSymbol{\nu}{\mathalpha}{letters}{"17}
\DeclareMathSymbol{\xi}{\mathalpha}{letters}{"18}
\DeclareMathSymbol{\pi}{\mathalpha}{letters}{"19}
\DeclareMathSymbol{\rho}{\mathalpha}{letters}{"1A}
\DeclareMathSymbol{\sigma}{\mathalpha}{letters}{"1B}
\DeclareMathSymbol{\tau}{\mathalpha}{letters}{"1C}
\DeclareMathSymbol{\upsilon}{\mathalpha}{letters}{"1D}
\DeclareMathSymbol{\phi}{\mathalpha}{letters}{"1E}
\DeclareMathSymbol{\chi}{\mathalpha}{letters}{"1F}
\DeclareMathSymbol{\psi}{\mathalpha}{letters}{"20}
\DeclareMathSymbol{\omega}{\mathalpha}{letters}{"21}
\DeclareMathSymbol{\varepsilon}{\mathalpha}{letters}{"22}
\DeclareMathSymbol{\vartheta}{\mathalpha}{letters}{"23}
\DeclareMathSymbol{\varpi}{\mathalpha}{letters}{"24}
\DeclareMathSymbol{\varrho}{\mathalpha}{letters}{"25}
\DeclareMathSymbol{\varsigma}{\mathalpha}{letters}{"26}
\DeclareMathSymbol{\varphi}{\mathalpha}{letters}{"27}
\DeclareMathSymbol{\Gamma}{\mathalpha}{letters}{"00}
\DeclareMathSymbol{\Delta}{\mathalpha}{letters}{"01}
\DeclareMathSymbol{\Theta}{\mathalpha}{letters}{"02}
\DeclareMathSymbol{\Lambda}{\mathalpha}{letters}{"03}
\DeclareMathSymbol{\Xi}{\mathalpha}{letters}{"04}
\DeclareMathSymbol{\Pi}{\mathalpha}{letters}{"05}
\DeclareMathSymbol{\Sigma}{\mathalpha}{letters}{"06}
\DeclareMathSymbol{\Upsilon}{\mathalpha}{letters}{"07}
\DeclareMathSymbol{\Phi}{\mathalpha}{letters}{"08}
\DeclareMathSymbol{\Psi}{\mathalpha}{letters}{"09}
\DeclareMathSymbol{\Omega}{\mathalpha}{letters}{"0A}

\begin{document}
\preprint{SAGA-HE-245-08}
\title{Meson mass at real and imaginary chemical potentials}

\author{Kouji Kashiwa}
\email[]{kashiwa@phys.kyushu-u.ac.jp}
\affiliation{Department of Physics, Graduate School of Sciences, Kyushu University,
             Fukuoka 812-8581, Japan}

\author{Masayuki Matsuzaki}
\email[]{matsuza@fukuoka-edu.ac.jp}
\affiliation{Department of Physics, Fukuoka University of Education, 
             Munakata, Fukuoka 811-4192, Japan}

\author{Hiroaki Kouno}
\email[]{kounoh@cc.saga-u.ac.jp}
\affiliation{Department of Physics, Saga University,
             Saga 840-8502, Japan}
             
\author{Yuji Sakai}
\email[]{sakai@phys.kyushu-u.ac.jp}
\affiliation{Department of Physics, Graduate School of Sciences, Kyushu University,
             Fukuoka 812-8581, Japan}             

\author{Masanobu Yahiro}
\email[]{yahiro@phys.kyushu-u.ac.jp}
\affiliation{Department of Physics, Graduate School of Sciences, Kyushu University,
             Fukuoka 812-8581, Japan}

\date{\today}

\begin{abstract}
Chemical-potential dependence of pi and sigma meson masses 
is analyzed at both real and imaginary chemical potentials, 
$\mu_\mathrm{R}$ and $\mu_\mathrm{I}$, 
by using the Polyakov-loop extended Nambu--Jona-Lasinio (PNJL) model 
that possesses both the extended ${\mathbb Z}_3$ symmetry and 
chiral symmetry. 
In the $\mu_\mathrm{I}$ region, 
the meson masses have the Roberge-Weiss periodicity. 
The $\mu_\mathrm{I}$ dependence of the meson masses becomes 
stronger as temperature increases. 
We argue that meson masses 
and physical quantities in the $\mu_\mathrm{R}$ region 
will be determined from lattice QCD data on meson masses in 
the $\mu_\mathrm{I}$ region by using the PNJL model, 
if the data are measured in the future. 
\end{abstract}

\maketitle

\section{Introduction}
Extensive studies for exploring 
the phase diagram of Quantum Chromodynamics
(QCD) have been done at finite temperature ($T$) and 
chemical potential ($\mu$).
In the study of the quark-gluon system, 
lattice QCD (LQCD) is a powerful method if $\mu=0$. 
LQCD, however, has the well known sign problem when 
the real part of $\mu$ is finite; 
for example, see Ref.~\cite{Kogut} and references therein.
Therefore, several approaches 
such as the reweighting method~\cite{Fodor}, 
the Taylor expansion method~\cite{Allton},
the analytic continuation  
from imaginary chemical potential $\mu_\mathrm{I}$
to real one $\mu_\mathrm{R}$~\cite{FP,Elia1,Elia2}
and so on 
are suggested 
to circumvent the difficulty, 
but those are still far from perfection.

Constructing the effective model is an approach complementary to 
the first-principle LQCD simulation. 
For example, the phase structure and light meson masses at finite $T$ 
and $\mu_\mathrm{R}$ are extensively investigated by the  
Nambu--Jona-Lasinio (NJL) model~\cite{AY,BR,Sca,Fuj,KKKN} and 
the the Polyakov-loop extended Nambu--Jona-Lasinio (PNJL) model
~\cite{Meisinger,Dumitru,Fukushima1,Ghos,Megias,Ratti1,Ciminale,Ratti2,Rossner,Hansen,Sasaki,Schaefer,Costa,Kashiwa1,Fu,Abuki,Hell,Tsai,Fukushima2}. 
The NJL model can treat the chiral symmetry breaking, but 
does not possess the confinement mechanism. 
Meanwhile, 
the PNJL model is designed \cite{Fukushima1} 
to treat the confinement mechanism approximately in addition to 
the symmetry breaking. In this sense, the PNJL model is 
superior to the NJL model.

In the PNJL model with two flavor quarks, 
the model parameters are usually determined from 
the pion mass and the 
pion decay constant at $T=\mu=0$ and 
LQCD data at finite $T$ and zero $\mu$; 
see Sec. II for the details. 
However, the strength of the vector-type four-point 
interaction can not be determined at zero $\mu$ and then remains 
as a free parameter, 
although the location of the critical endpoint of the chiral phase transition 
in the $\mu_\mathrm{R}$ region is found 
to be sensitive to the strength~\cite{KKKN,Kashiwa2,Fukushima2}. 
Thus, it is highly nontrivial how large the strength of the vector-type 
interaction is and whether the PNJL model well simulates the 
$\mu$ dependence of the phase structure and light meson masses. 
This should be tested directly from QCD. 
Fortunately, this is possible in the $\mu_\mathrm{I}$ region, 
since LQCD is feasible there 
because of the absence of the sign problem. 
Furthermore, it is possible to determine the strength 
of the interaction
,for example the vector-type four quark and 
the higher-order multi-quark interaction, 
from 
LQCD data in the $\mu_\mathrm{I}$ region, as proposed by 
our previous paper~\cite{Sakai}.

In addition, the canonical partition function $Z_\mathrm{C}(n)$ 
with real quark number $n$ can be obtained 
as the Fourier transform of the grand-canonical one 
$Z_\mathrm{GC}(\theta)$ with $\mu=i\mu_\mathrm{I}=i \theta T$~\cite{RW}: 
\bea
Z_\mathrm{C}(n)=\frac{1}{2\pi}\int_{-\pi}^{\pi} d\theta e^{-in\theta}Z_\mathrm{GC}(\theta) .
\label{trans-to-n}
\eea 
Thus, the thermodynamic potential, 
$\Omega_{\rm QCD}(\theta)=-T \ln(Z_\mathrm{GC}(\theta))$, 
at finite $\theta$ 
contains the QCD dynamics 
at real $n$ and hence at finite $\mu_\mathrm{R}$ in principle. 
Therefore, we can confirm the reliability of the PNJL model in the 
$\mu_\mathrm{R}$ region by comparing the model results with 
LQCD ones in the $\mu_\mathrm{I}$ region.

Roberge and Weiss (RW) found that $\Omega_{\rm QCD}(\theta)$ 
has a periodicity, 
$\Omega_{\rm QCD}(\theta)=\Omega_{\rm QCD}(\theta+2\pi k/3)$, 
in the $\mu_\mathrm{I}$ region \cite{RW}, 
where $k$ is any integer. 
The RW periodicity indicates 
that QCD is invariant under the extended ${\mathbb Z}_3$ 
transformation, that is, 
the combination of the ${\mathbb Z}_3$ transformation and 
the coordinate transformation $\theta \to \theta + 2\pi/3$, 
as shown later in Sec. \ref{PNJL}; 
see our previous works~\cite{Sakai} for the details. 
At the present stage the PNJL model is only 
a realistic model that possesses both 
the extended ${\mathbb Z}_3$ symmetry and chiral symmetry. 
The PNJL model results are then consistent with LQCD ones 
particularly in the $\theta$-dependence of the Polyakov loop, 
the quark number density and the chiral condensate~\cite{Sakai}. 
In the PNJL model, 
we do not need any extrapolation from 
the $\mu_\mathrm{I}$ to the $\mu_\mathrm{R}$ region, 
since the model calculation 
is feasible at finite $\mu_\mathrm{R}$ 
with the input parameters 
determined so as to reproduce LQCD data 
in the $\mu_\mathrm{I}$ region. 
We call this procedure the imaginary chemical-potential matching approach 
(the $\theta$-matching approach) 
in this paper.

In this paper, using the PNJL model, 
we predict the $\mu$ dependence of pi and sigma meson masses 
in the real and imaginary $\mu$ regions, and 
argue that meson masses 
and physical quantities in the $\mu_\mathrm{R}$ region 
will be determined from LQCD data on meson masses 
in the $\mu_\mathrm{I}$ region by using the PNJL model, 
if the data are measured in the future.
Concretely, the following four points are argued. 
First, we show that the meson masses have the RW periodicity. 
Second, the $\theta$ dependence of meson masses is found to become large as 
$T$ increases. 
We then recommend that LQCD simulations on meson masses 
in the $\mu_\mathrm{I}$ region be made at 
higher $T$ near the pseudo-critical 
temperature $T_{c}$ of the deconfinement phase transition 
at $\mu=0$. 
Third, the validity of two extrapolations 
from $\mu=\mu_\mathrm{I}$ to 
$\mu=\mu_\mathrm{R}$ is discussed by comparing results of the 
extrapolations with the PNJL one. 
In recent LQCD calculations at finite $\theta$, 
only small lattice sizes are taken, 
so that pion mass evaluated in Ref.~\cite{Wu} 
is unnaturally large. 
This indicates that the bare quark mass ($m_0$) taken there is 
rather large. 
Finally, we discuss how sensitive meson masses are to the value of $m_0$, 
comparing two cases of $m_0=5.5$ and $80$~MeV. 
This sort of model prediction is important 
before doing heavy LQCD calculations with large lattice size 
in the $\mu_\mathrm{I}$ region.

We briefly explain the PNJL model in section II 
and present equations for meson masses in section III. 
In section IV, numerical results are shown on the $\mu$-dependence of 
pi and sigma meson masses, and the validity 
of two extrapolation methods is discussed. 
Section V is devoted to summary.

\section{PNJL model}
\label{PNJL}
In this section, 
we briefly review the PNJL model; 
see \cite{Sakai} for the details. 
The Lagrangian density of the two-flavor PNJL model is
\begin{align}
 {\cal L}_{\rm PNJL}  
=& {\bar q}(i \gamma_\nu D^\nu -m_0)q  + G_{\rm s}[({\bar q}q )^2 
  +({\bar q }i\gamma_5 {\vec \tau}q )^2]
\nonumber\\
 & -{\cal U}(\Phi [A],{\bar \Phi} [A],T) , 
\label{L}
\end{align} 
where $D^\nu=\partial^\nu + iA^\nu$, 
$q$ denotes the quark field with two flavor and 
the current quark mass $m_0$.  
The field $A^\nu$ is defined as 
$A^\nu=\delta^{\nu}_{0}gA^0_a{\lambda_a\over{2}}$
with the gauge field $A^\nu_a$, 
the Gell-Mann matrix $\lambda_a$ and 
the gauge coupling $g$.
The matrix ${\vec \tau}$ stands for the isospin matrix 
and $G_{\rm s}$ denotes the coupling constant of the scalar-type 
four-quark interaction. 
For simplicity, we neglect the vector-type four-quark interaction, 
since it does not change the conclusion of this paper qualitatively.

In the PNJL model, the gauge field $A_\mu$ is treated 
as a homogeneous and static background field $A_0$, that is, 
$A_\mu =\delta_{0\mu}A_0$. 
In the Polyakov gauge, 
the Polyakov loop and its Hermitian conjugate, 
$\Phi$ and ${\bar \Phi}$, are diagonal in color space:
\bea
\Phi &=&{1\over{3}}{\rm tr}_{\rm c}(L),
~~~~~\bar{\Phi} ={1\over{3}}{\rm tr}_{\rm c}({\bar L})
\label{Polyakov}
\eea 
with 
\bea
L  &=& e^{i A_4/T}={\rm diag} \Bigl(
e^{i \phi_a/T},e^{i \phi_b/T},e^{-i (\phi_a+\phi_b)/T} 
\Bigr) ,
\eea
where $\phi_a$ and $\phi_b$ are classical variables and $A_4=iA_0$.

We make the mean field approximation (MFA) to the quark-quark interactions in 
\eqref{L}, as follows. In \eqref{L}, the operator product ${\bar q}q$ is 
first divided into $ {\bar q}q = \sigma  + ({\bar q}q)'$ with 
the mean field $\sigma \equiv \langle {\bar q}q \rangle$ 
and the fluctuation $({\bar q}q)'$. 
Ignoring the higher-order terms of $({\bar q}q)'$ 
in the rewritten Lagrangian and resubstituting 
$ ({\bar q}q)' = {\bar q}q -\sigma $ into the approximated Lagrangian, 
one can obtain a linearized Lagrangian based on MFA, 
\begin{align}
 {\cal L}_{\rm PNJL}^{\rm MFA}  
=& {\bar q}(i \gamma_\nu \partial^\nu+i\gamma_0A_4 -M)q  - G_{\rm s}\sigma^2 
\nonumber\\
 & -{\cal U}(\Phi [A],{\bar \Phi} [A],T) ,
\label{linear-L}
\end{align} 
where $M$ is the effective quark mass defined by 
$M=m_0-2G_{\rm s}\sigma$. 
In \eqref{linear-L}, 
use has been made of $\langle {\bar q} i \gamma_5 {\vec \tau} q \rangle=0$, 
because the ground state is assumed to be invariant under 
the parity transformation. In the MFA Lagrangian 
${\cal L}_{\rm PNJL}^{\rm MFA}$, quark fields 
interact only 
with the homogeneous and static back ground fields $A_0$ and $\sigma$.  
Hence, we can easily make the path integral over the quark field to 
get the thermodynamic potential per unit volume, 
\begin{align}
&\Omega_{\rm PNJL} 
= -2 N_{\rm f} \int \frac{d^3 p}{(2\pi)^3}
   \Bigl[ 3 E_p \notag \\
&+ \frac{1}{\beta}
           \ln~ [1 + 3(\Phi+{\bar \Phi} e^{-\beta (E_p-\mu )}) 
           e^{-\beta (E_p-\mu)}+ e^{-3\beta (E_p-\mu)}] \notag\\
&+ \frac{1}{\beta} 
           \ln~ [1 + 3({\bar \Phi}+{\Phi e^{-\beta (E_p+\mu)}}) 
              e^{-\beta (E_p+\mu)}+ e^{-3\beta (E_p+\mu)}]
	      \Bigl]
	      \nonumber\\
&+U_{\rm M}+{\cal U},
\label{PNJL-Omega}
\end{align}
where $\mu=\mu_\mathrm{R}+i\mu_\mathrm{I}=\mu_\mathrm{R}+iT\theta$, 
$E_p=\sqrt{{\bf p}^2+M^2}$ and  
$U_{\rm M}= G_{\rm s} \sigma^2$.

The thermodynamic potential $\Omega_{\rm PNJL}$ is 
invariant under the extended ${\mathbb Z}_3$ transformation~\cite{Sakai}, 
\begin{align}
&e^{\pm i \theta} \to e^{\pm i \theta} e^{\pm i{2\pi k\over{3}}},\quad  
 \Phi(\theta)  \to \Phi(\theta) e^{-i{2\pi k\over{3}}}, 
\nonumber\\
&{\bar \Phi}(\theta) \to {\bar \Phi}(\theta) e^{i{2\pi k\over{3}}} .
\label{eq:K2}
\end{align}
This is easily understood by introducing the modified Polyakov loop, 
$\Psi \equiv e^{i\theta}\Phi$ and 
${\bar \Psi} \equiv e^{-i\theta}{\bar \Phi}$,
invariant under the the extended ${\mathbb Z}_3$ transformation (\ref{eq:K2}), 
since $\Omega_{\rm PNJL}$ is described as a function of only 
the extended ${\mathbb Z}_3$ invariant quantities, 
$\Psi$, ${\bar \Psi}$, $\sigma$ and 
$e^{3i\theta}$
\begin{align}
\Omega_{\rm PNJL}  = & -2 N_f \int \frac{d^3{\rm p}}{(2\pi)^3}
          \Bigl[ 3 E_p 
          + \frac{1}{\beta}\ln~ [1 + 3\Psi e^{-\beta E_p}
\notag\\
          &+ 3\Psi^{*}e^{-2\beta E_p}e^{i 3 \theta}
          + e^{-3\beta E_p}e^{i 3 \theta}]
\notag\\
          &+ \frac{1}{\beta} 
           \ln~ [1 + 3\Psi^{*} e^{-\beta E_p}
          + 3\Psi e^{-2\beta E_p}e^{-i 3 \theta}
\notag\\
          &+ e^{-3\beta E_p}e^{-i 3 \theta}]
	      \Bigl]+U_{\rm M}+ {\cal U} .
\end{align}
The physical quantities $X=\sigma, \Psi$ and ${\bar \Psi}$ 
are determined by the stationary conditions 
$\partial \Omega_{\rm PNJL}/\partial X=0$. 
These equations include the $\theta$-dependence only through 
the factor $e^{3i\theta}$, 
indicating that the $X$ have 
the RW periodicity, $X(\theta)=X(\theta+2\pi k/3)$. 
Inserting the solutions back to $\Omega_{\rm PNJL}$, 
one can see that $\Omega_{\rm PNJL}$ also has the RW periodicity, 
$\Omega_{\rm PNJL}(\theta)=\Omega_{\rm PNJL}(\theta+2\pi k/3)$; 
see \cite{Sakai} for the details.

We take the three-dimensional momentum cutoff
because this model is nonrenormalizable, 
\begin{equation}
\int \frac{d^3 p}{(2 \pi)^3}\to 
{1\over{2\pi^2}} \int_0^\Lambda dp p^2.
\end{equation}
Hence, the present model has three parameters 
$m_0$, $\Lambda$ and $G_{\rm s}$.  
We take $\Lambda =0.6315$ GeV and $G_{\rm s}=5.498$ GeV$^{-2}$ so 
as to reproduce the pion decay constant $f_{\pi}=93.3$ MeV and 
the pion mass $M_{\pi}=138$ MeV at $T=\mu=0$, when 
a realistic quark mass $m_0=5.5$ MeV is taken \cite{Kashiwa2,Kashiwa3}.

We use ${\cal U}$ of Ref.~\cite{Ratti1} fitted 
to LQCD data in the pure gauge theory 
at finite $T$~\cite{Boyd,Kaczmarek}: 
\begin{align}
&{{\cal U}\over{T^4}} =  -\frac{b_2(T)}{2} {\bar \Phi}\Phi
              -\frac{b_3}{6}({\bar \Phi}^3+ \Phi^3)
              +\frac{b_4}{4}({\bar \Phi}\Phi)^2, \label{eq:E13} \\
&b_2(T)   = a_0 + a_1\Bigl(\frac{T_0}{T}\Bigr)
                 + a_2\Bigl(\frac{T_0}{T}\Bigr)^2
                 + a_3\Bigl(\frac{T_0}{T}\Bigr)^3.  \label{eq:E14} 
\end{align}
Following Ref.~\cite{Sakai}, we take $T_0=190$ MeV so as to 
reproduce the pseudo-critical 
temperature $T_{c}$ of the deconfinement phase transition 
at $\mu=0$ evaluated 
by LQCD; specifically, $T_{c}$ is 170~MeV in 
the PNJL model, whereas it is $173 \pm 8$~MeV in full LQCD calculations~\cite{Karsch}. 
Thus, the present four parameters are determined from the pion mass and the 
pion decay constant at $T=\mu=0$ and LQCD data at $T >0$ and $\mu=0$. 
However, if the vector-type four-quark interaction 
$({\bar q }\gamma_{\mu} q)^2$ 
is added to $\cal L$, the strength 
can not be determined at $\mu=0$, since its mean field 
$n=\langle {\bar q} \gamma_{0} q \rangle$ is zero there. 
The strength will be 
determined by physical quantities at finite $\theta$ such as meson masses, 
if they become available in the future.

\section{Meson mass}

In this section, we consider pion and sigma meson and 
derive equations for the meson masses, following Ref~\cite{Hansen}. 
Correlators of current operators carry physical mesons with the 
quantum number. 
The pseudoscalar isovector current with the same quantum number 
as pion is 
\beq
  {J_P}^a(x) = \bar q(x) i \gamma_5 \tau^a q(x)\quad 
\eeq
and the scalar isoscalar current with the same quantum number 
as sigma meson is 
\beq
  {J_S}(x) = \bar q(x) q(x) - \langle\bar q(x) q(x) \rangle .
\eeq
The Fourier transform of the mesonic correlation function 
$\langle 0 | T \left( J^a_\xi(x) J^{b\dagger}_\xi(0) \right) | 0 \rangle$
is defined as
\begin{align}
C_{\xi\xi}^{ab} (q^2) 
&\equiv  i \int d^4x e^{i q.x}
 \langle 0 | T \left( J^a_\xi(x) J^{b\dagger}_\xi(0) \right)  
 | 0 \rangle  \nonumber\\
&= C_{\xi\xi}(q^2) \delta^{ab} ,
\end{align}
where $\xi=P$ for pion and $S$ for sigma meson and $T$ stands for 
the time-ordered product. 
Using the random phase approximation (the ring approximation),
one can obtain the Schwinger-Dyson equation for $C_{\xi\xi}$ 
at $T=\mu=0$ where $\Phi={\bar \Phi}=0$,    
\bea
  C_{\xi\xi}(q^2) &=& \Pi_{\xi\xi}(q^2) +2G_{\rm s} \Pi_{\xi\xi} (q^2) C_{\xi\xi}
  \label{SD-eq}
\eea
with the one-loop polarization function 
\bea
  \Pi_{\xi\xi} & \equiv & (-i) \int \frac{d^4 p}{(2\pi)^4} 
  {\rm Tr} \left( \Gamma_\xi iS(p+q) \Gamma_\xi iS(q) \right) ,
\eea
where $S(q)$ is the quark propagator in the Hartree approximation and 
$\Gamma_\xi=\Gamma^a_{P}=i\gamma_5 \tau^a$ for pion and 
$\Gamma_\xi=\Gamma_{S}= 1$ for sigma meson. 
In the random phase approximation, the solution to \eqref{SD-eq} is given as 
\bea
C_{\xi\xi} &=& \frac{\Pi_{\xi\xi}(q^2)}{1 - 2G_{\rm s} \Pi_{\xi\xi}(q^2)} . 
\eea
Noting that meson mass $M_{\xi}$ ($\xi=\pi$ and $\sigma$) is a pole mass of $C_{\xi\xi}(q^2)$ and 
taking the rest frame $q=(q_0,0)$, one can get an equation for $M_{\xi}$ as 
\begin{align}
\big[1 - 2G_{\xi\xi} \Pi_{\xi\xi}(q_0)\big]\big|_{q_0=M_\xi}=0.   
\label{mmf}
\end{align}
The explicit forms of 
$\Pi_{\rm PP} (q_0)$ and $\Pi_{\rm SS} (q_0)$ are given as 
\begin{align}
&\Pi_{\rm PP} (q_0) 
\nonumber\\
&= -i {\rm Tr} \int \frac{d^4p}{(2 \pi)^4} 
       [i\gamma_5 {\vec \tau} iS(p+q_0/2)
        i\gamma_5 {\vec \tau} iS(p-q_0/2)] 
\nonumber \\
&=  4  i~{\rm tr_c} \int \frac{d^4p}{(2 \pi)^4} 
    \frac{(p+q_0/2)(p-q_0/2)-M^2}{[(p+q_0/2)^2-M^2+i\epsilon]}
\nonumber\\
&   \hspace{1cm} \times \frac{1}{[(p-q_0/2)^{2}-M^2+i\epsilon]},\\
&\Pi_{\rm SS} (q_0) 
\nonumber\\
&= -i {\rm Tr} \int \frac{d^4p}{(2 \pi)^4} 
       [iS(p+q_0/2)iS(p-q_0/2)]
\nonumber \\
&=  4 i~{\rm tr_c} \int \frac{d^4p}{(2 \pi)^4} 
    \frac{(p+q_0/2)(p-q_0/2)+M^2}{[(p+q_0/2)^2-M^2+i\epsilon]}
\nonumber\\
&   \hspace{1cm} \times \frac{1}{[(p-q_0/2)^{2}+M^2+i\epsilon]},
\end{align}
where ${\rm Tr}$ denotes 
${\rm tr}_{\rm Dirac} \otimes {\rm tr}_{\rm Spin} \otimes 
 {\rm tr}_{\rm c} \otimes {\rm tr}_{\rm Flavor}$. 

When $T$ and $\mu$ are finite, the corresponding equations are 
obtained by the replacement 
\begin{align}
&p_0 \to i \omega_n + \mu + iA_4= (2n+1) \pi T +\mu + iA_4, 
\nonumber\\
&\int \frac{d^4p}{(2 \pi)^4} 
\to iT\sum_n \int \frac{d^3p}{(2 \pi)^3}. 
\end{align}
Also in this case, an equation for meson mass $M_\xi$ is of the same form 
as \eqref{mmf}, but 
the polarization functions are obtained in more complicated forms: 
\begin{align}
\Pi_{\rm PP} (q_0) 
&= -2 N_{\rm f}[2 A(\mu) -q_0^2 B(q_0,\mu)]
\label{polp}
\end{align}
for pion  and 
\begin{align}
\Pi_{\rm SS} (q_0) 
&= -2 N_{\rm f}[2 A(\mu) -(q_0^2-4M^2) B(q_0,\mu)],
\label{pols}
\end{align}
for sigma meson. Here, 
$A(\mu)$ and $B(q_0,\mu)$ denote loop integrals~\cite{Fetter} defined 
by 
\begin{align}
A(\mu) &= \int \frac{d^3 p}{(2\pi)^3} 
            \frac{1-f_{\rm PNJL}(E_p-\mu)-f_{\rm PNJL}(E_p+\mu)}{2 E_p},
\label{oli1}
\\
B(q_0,\mu) &= \int \frac{d^3 p}{(2\pi)^3} 
        \frac{1-f_{\rm PNJL}(E_p-\mu)-f_{\rm PNJL}(E_p+\mu)}{E_p(q_0^2-4E_p)}, 
\label{oli2}
\end{align}
where $f_{\rm PNJL}(E_p\pm\mu)$  is 
the modified Fermi-Dirac distribution function~\cite{Hansen} defined by 
\begin{align}
&f_{\rm PNJL} (E_p - \mu)  
\nonumber\\
&=\frac{(\Phi+2{\bar \Phi}e^{-\beta(E_p - \mu)})e^{-\beta(E_p - \mu)}
             +e^{-3\beta(E_p - \mu)}}
       {1+3(\Phi+{\bar \Phi}e^{-\beta(E_p - \mu)})e^{-\beta(E_p - \mu)}
             +e^{-3\beta(E_p - \mu)}} .     
\label{FD}
\end{align} 
The distribution function $f_{\rm PNJL} (E_p + \mu) $
is obtained by replacing $-\mu \to +\mu$, $\Phi \to {\bar \Phi}$ and ${\bar \Phi} \to \Phi$.
In actual PNJL calculations, the imaginary part of 
meson mass $M_{\xi}$ is assumed to be negligible in \eqref{mmf}; 
this assumption is exactly satisfied when $M_x$ is smaller 
than twice the dynamical quark mass, 
$2M$, and approximately satisfied for $M_x$ slightly above $2M$.

In the case of $\mu=i\mu_\mathrm{I}=iT\theta$,
the corresponding distribution functions  
\begin{align}
&f_{\rm PNJL} (E_p - iT \theta) 
\nonumber\\ 
&=\frac{ \Psi+2{\bar \Psi}e^{-\beta E_p}e^{3i\theta}
             +e^{-3\beta E_p}e^{3i\theta}}
       {1+3(\Psi+{\bar \Psi}e^{-\beta E_p}e^{3i\theta})
             +e^{-3\beta E_p}e^{3i\theta}} ,             
\label{FD2}
\end{align}
depend only on the extended ${\mathbb Z}_3$ invariant quantities, 
$\Psi$, ${\bar \Psi}$, $\sigma$ and $e^{3i\theta}$. 
The distribution function $f_{\rm PNJL} (E_p - iT \theta)$
is obtained by replacing $\theta \to -\theta$, $\Psi \to {\bar \Psi}$ and ${\bar \Psi} \to \Psi$.
Therefore, the distribution functions are also extended ${\mathbb Z}_3$ 
invariant, and hence they have the RW periodicity. 
Furthermore, the polarization functions $\Pi_{\xi\xi}$ 
depend on $\theta$ only through 
the modified Fermi-Dirac distribution function 
$f_{\rm PNJL}(E_p \pm i T \theta)$. 
Therefore, the meson masses have the RW periodicity.

\section{Numerical Results}

\begin{figure}[htbp]
\begin{center}
 \includegraphics[width=0.4\textwidth]{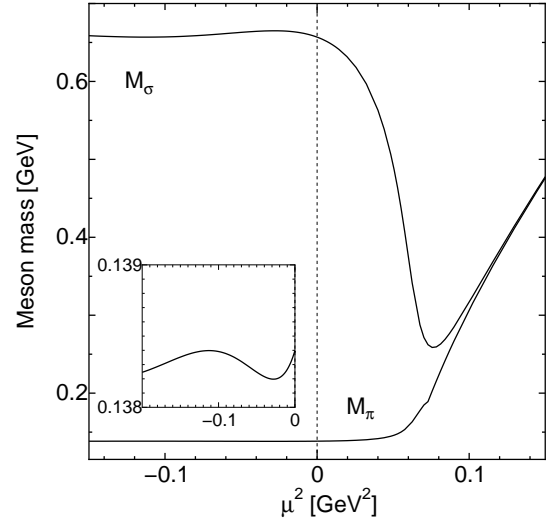}
\end{center}
\caption{The $\mu^2$-dependence of 
sigma and pi meson masses, 
$M_{\sigma}$ and $M_{\pi}$, at $T=160$ MeV. 
The inset represents the pion mass near $\mu=0$. 
}
\label{Fig-1}
\end{figure}

\begin{figure}[htbp]
\begin{center}
 \includegraphics[width=0.4\textwidth]{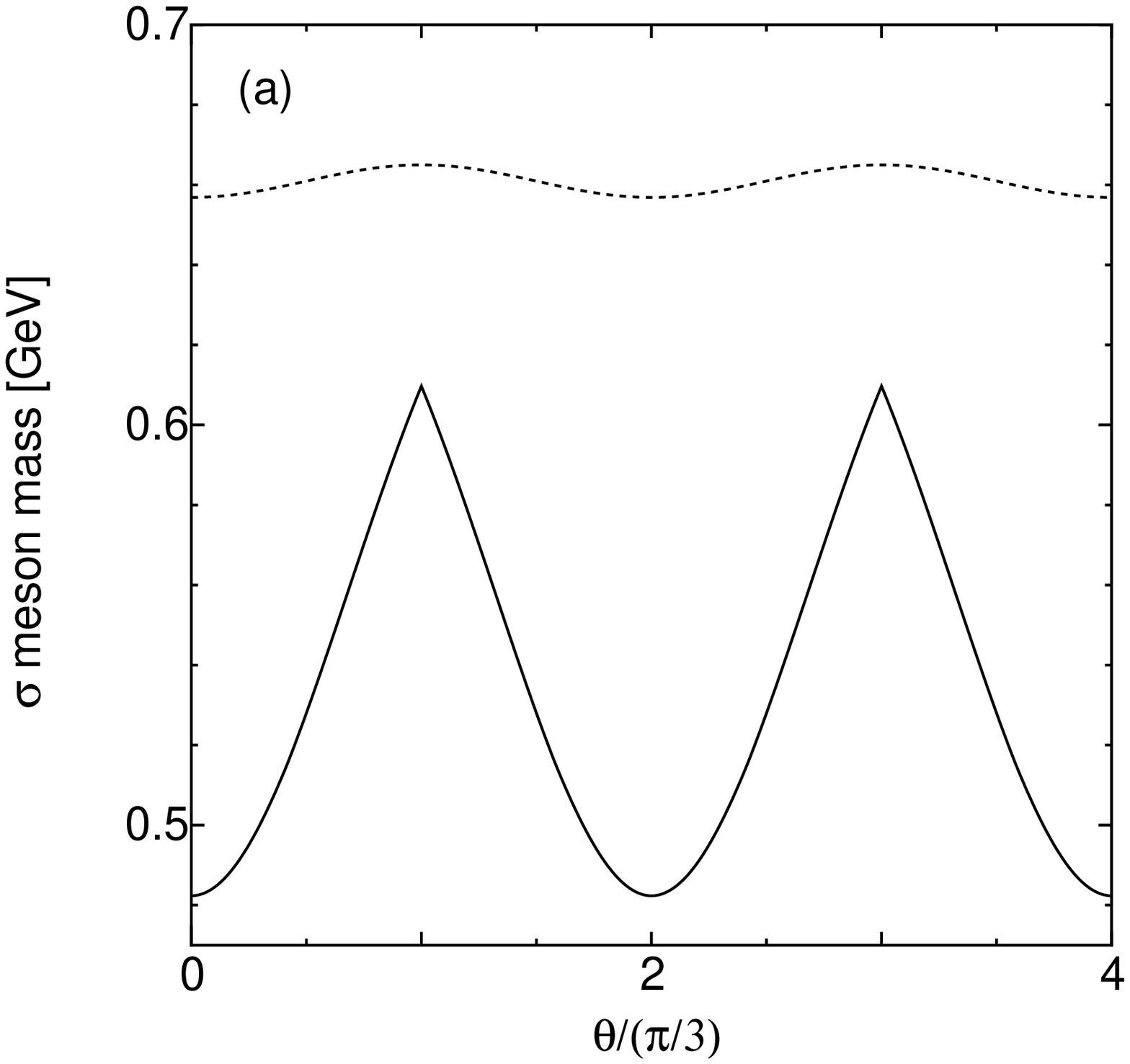} 
 \includegraphics[width=0.4\textwidth]{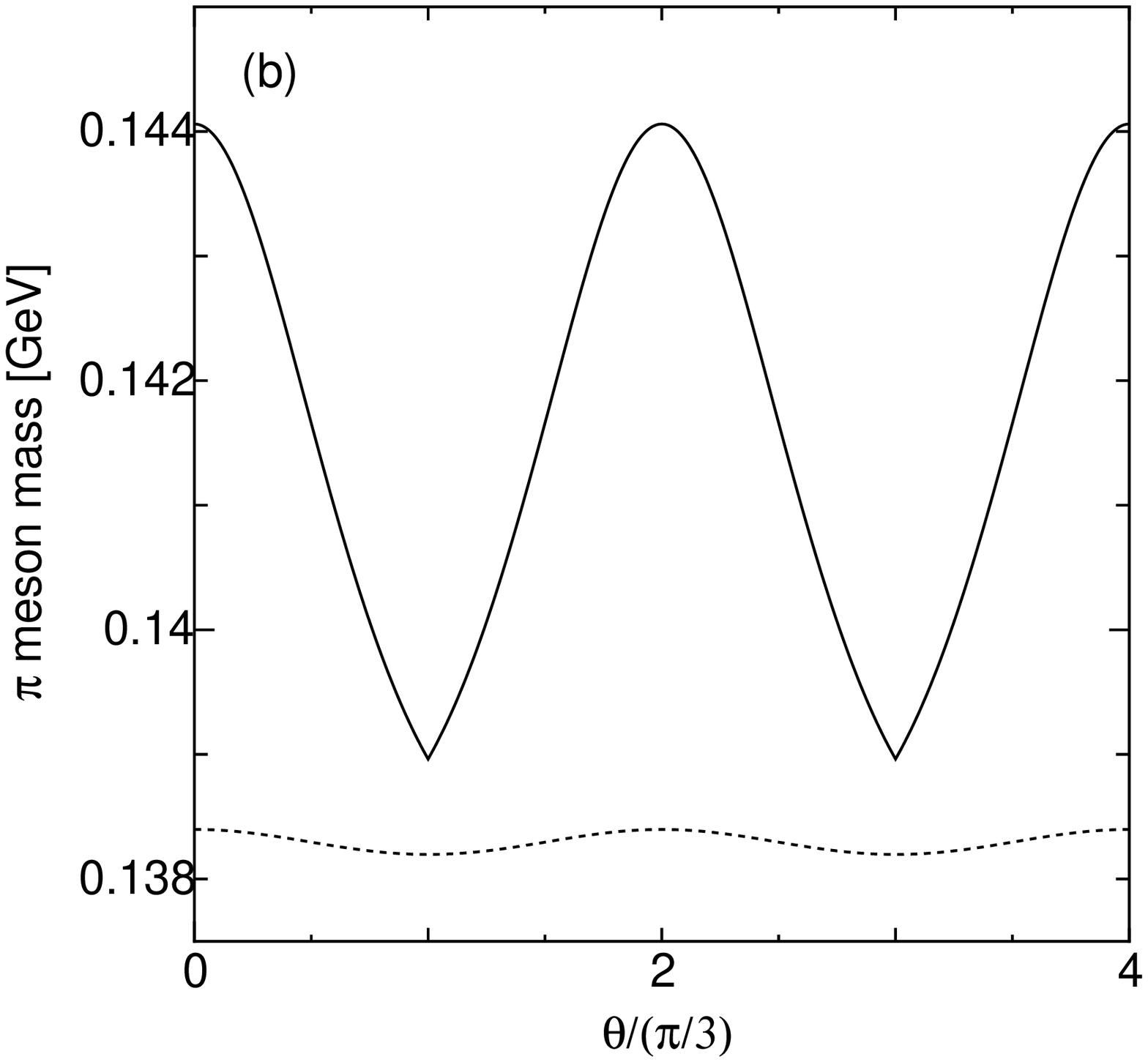}
\end{center}
\caption{ $\theta$ dependence of 
(a) sigma and (b) pi meson masses.
Dotted curves denote the results of $T=160$ MeV and
solid curves correspond to the results of $T=200$ MeV.
Note that the scales of the vertical axes are different between 
panels (a) and (b).}
\label{Fig-2}
\end{figure}

First, we investigate the $\mu$ dependence of pi and sigma meson masses, 
$M_{\pi}$ and $M_{\sigma}$, in the $\mu_\mathrm{R}$ and $\mu_\mathrm{I}$ 
regions. 
It is found from \eqref{oli1} and \eqref{oli2} that 
 $M_{\pi}$ and $M_{\sigma}$ are symmetric under the interchange 
$\mu\leftrightarrow-\mu$, indicating that they are functions of $\mu^2$. 
Figure~\ref{Fig-1} presents the $\mu^2$ dependence of 
$M_{\sigma}$ and $M_{\pi}$ in the case of $T=160$~MeV that is near 
the pseudo-critical temperature $T_c=170$~MeV of 
the deconfinement phase transition at $\mu=0$. 
They are smooth at $\mu^2=0$, as expected. 
This makes it possible the analytic continuation 
of meson masses, $M_{\pi}(\mu)$ and $M_{\sigma}(\mu)$, 
from $\mu=i\mu_\mathrm{I}$ to $\mu=\mu_\mathrm{R}$. 
This is important for LQCD simulation, 
although the PNJL calculation does not need the analytic continuation. 
In the right-half panel representing the $\mu_\mathrm{R}$ region,  
$M_{\pi}$ and $M_{\sigma}$ agree with each other when $\mu^2 \greata $0.1 GeV$^2$. 
This clearly exhibits that the restoration of chiral symmetry takes place 
at $\mu^2 \greata $0.1 GeV$^2$~\cite{Hansen}. 
In the left-half panel representing the $\mu_\mathrm{I}$ region, 
$M_{\pi}$ and $M_{\sigma}$ look almost constant in this scale, but 
one can see oscillations of the masses 
on closer inspection shown by the inset  and 
Fig.~\ref{Fig-2} 
that presents $M_{\pi}$ and $M_{\sigma}$ as a function of $\theta$.

In our previous paper~\cite{Sakai}, we showed that 
$\theta$-odd quantities with 
the RW periodicity such as $d\Omega_{\rm PNJL}/d\theta$ and 
the imaginary part of $\Psi$ are discontinuous 
at $\theta=\pi/3$ (mod $2\pi/3$) and $T \ge 190$~MeV.
This first-order transition is called the RW phase transition. 
Meanwhile, $\theta$-even quantities with 
the RW periodicity such as $\sigma$ and 
the real part of $\Psi$ are discontinuous in their derivative~\cite{Sakai}.  
This indicates that the meson masses $M_{\xi}$ have 
the same 
property as the chiral condensate. Actually, Fig.~\ref{Fig-2} shows that 
the $M_{\xi}$ are $\theta$-even functions with the RW periodicity and then 
not smooth on the RW phase transition line.

As an important property, 
the phase of oscillation is opposite 
to each other between $M_\sigma$ and $M_\pi$. 
This can be understood from their slopes at 
$\mu^2=0$; $dM_\sigma/d\mu^2<0$ while $dM_\pi/d\mu^2>0$ 
as shown in the inset of Fig.~\ref{Fig-1}. 
These slopes reflect the 
chiral symmetry restoration appearing at $\mu^2>0$. 
Another point to be noted is the difference in their amplitudes; 
the amplitude of oscillation is relatively larger in $M_\sigma$ 
whereas very small in $M_\pi$. 
This can be understood from (\ref{polp}) 
and (\ref{pols}); the latter has a term depending on the effective 
quark mass $M$ 
and consequently reflects the chiral symmetry restoration directly, 
while the former does not.

The meson mass $M_{\xi}(\theta)$ is a $\theta$-even function with the RW 
periodicity in the $\mu_\mathrm{I}$ region. 
This means that $M_{\xi}(\theta)$ can be expanded in terms of 
$\cos(3k\theta)$ with integer $k$: 
\bea
M_{\xi}(\theta)=\sum_{k=0} a_k(T) \cos{(3k\theta)}.
\label{expansion} 
\eea
When $T=160$~MeV, the normalized coefficients $a_k(T)/a_0(T)$ 
are about 0.1~\% for $k=1$ and negligibly small for $k \ge 2$. 
The neglect of the normalized coefficients with $k \ge 2$ 
is a good approximation at $T$ near and below $T_{\rm c}$. 
Particularly in the strong coupling limit of LQCD, 
the $a_k$ with $k \ge 2$ are known to be zero~\cite{Miura}. 
Hence, $M_{\xi}(\theta)$ is approximated into  
\begin{align}
M_{\xi}^{\mathrm{fit}} 
&= A(T) \cos(3\theta) + C(T),
\label{fit}
\end{align}
where $A(T)$ and $C(T)$ are determined from 
$M_{\xi}$ at $\theta=0$ and $\pi/3$ as 
\begin{align}
A(T)=[ M_{\xi}(T,\theta=0)-M_{\xi}(T,\theta=\pi/3) ]/2,\\
C(T)=[ M_{\xi}(T,\theta=0)+M_{\xi}(T,\theta=\pi/3) ]/2 .
\end{align}

\begin{figure}[htbp]
\begin{center}
 \includegraphics[width=0.4\textwidth]{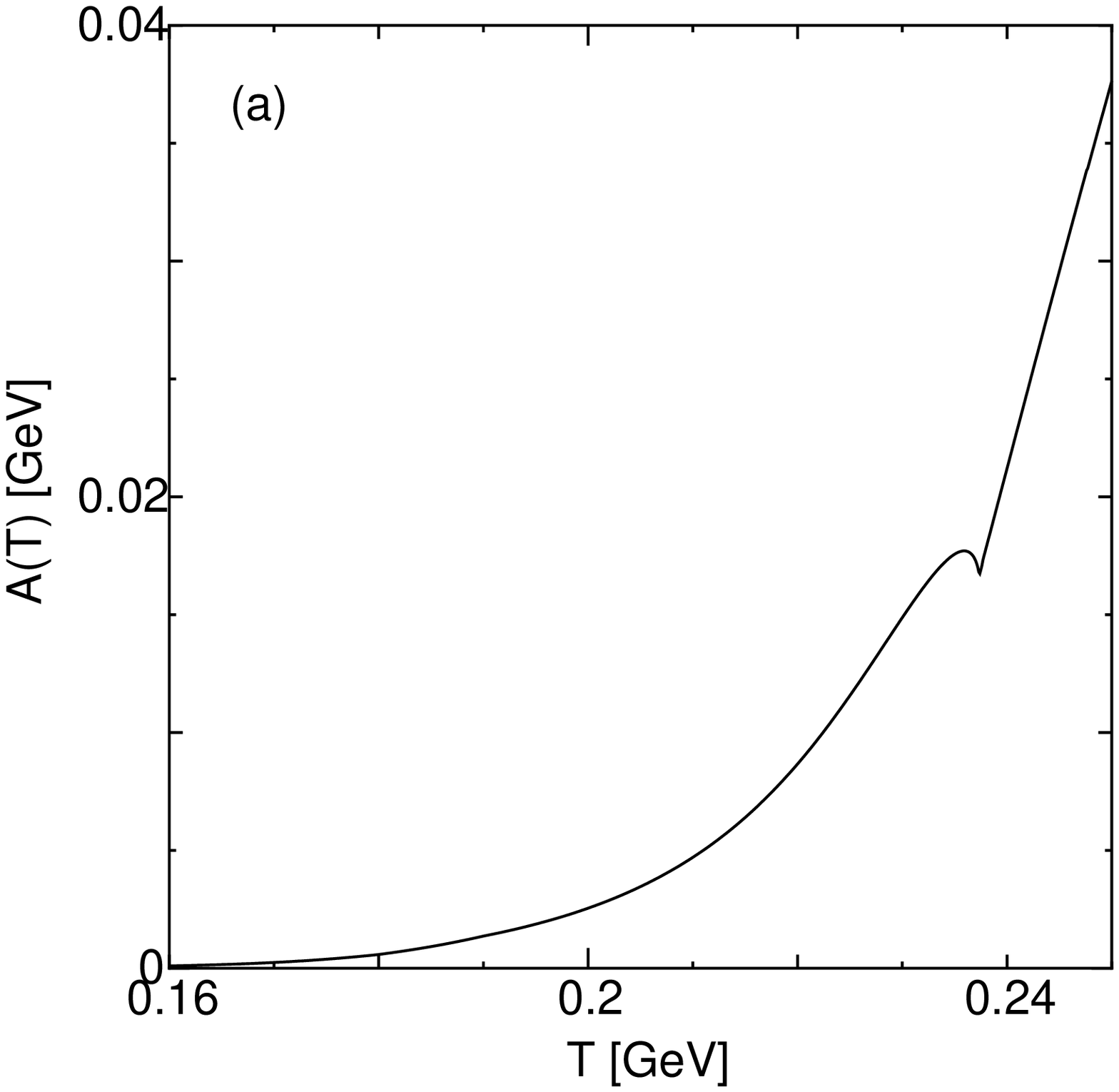}
 \includegraphics[width=0.4\textwidth]{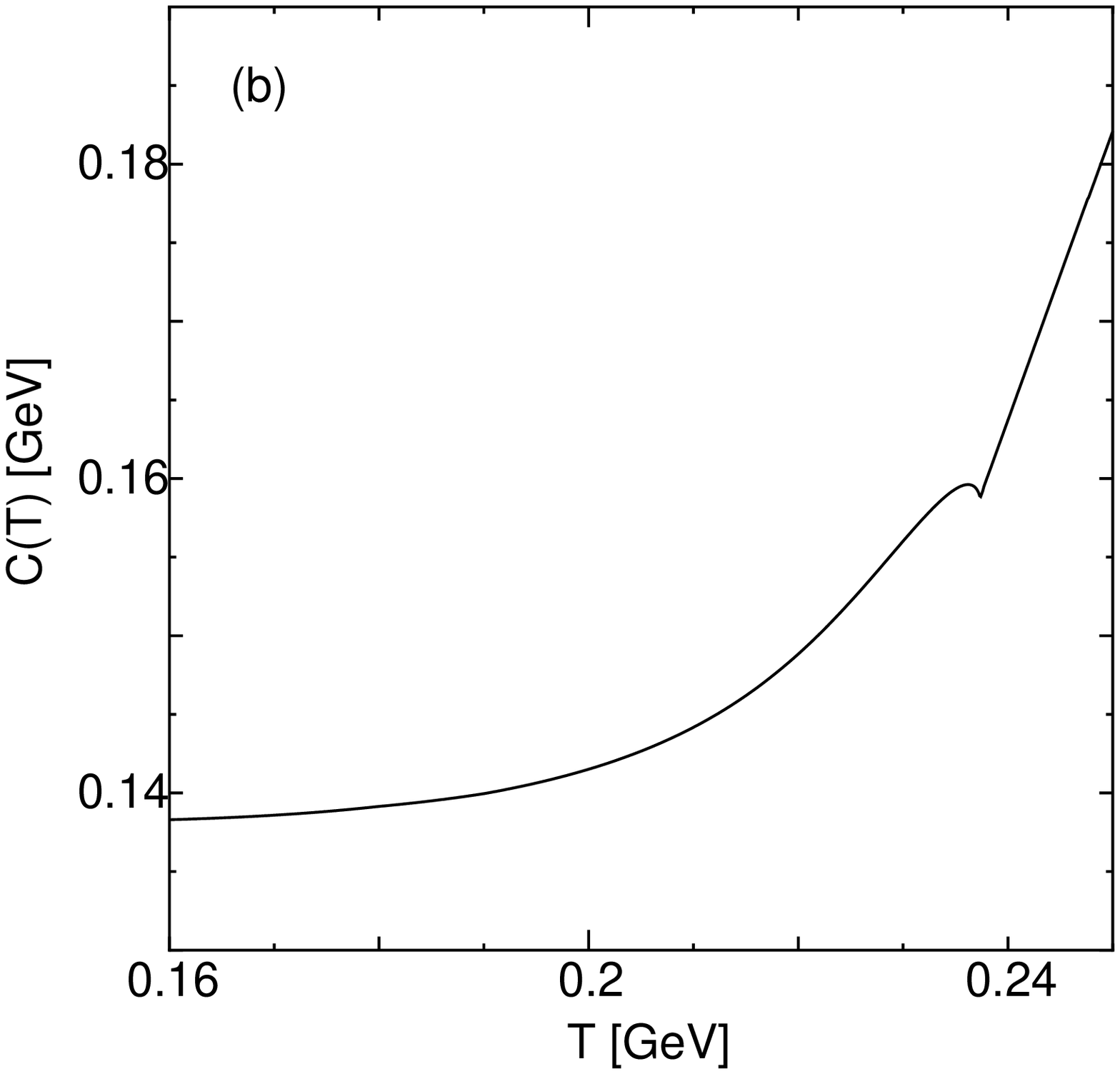}
\end{center}
\caption{$T$-dependence of $A(T)$ and $C(T)$ for the case of pion mass.}
\label{Fig-4}
\end{figure}

Figure~\ref{Fig-4} shows $A$ and $C$ for the case of $M_{\pi}$. 
They smoothly increase 
with increase in $T$ except a dip around $T=237$~MeV that 
comes from a threshold effect due to 
$\pi \rightarrow {\rm quark} + {\rm antiquark}$. 
The ratio $A/C$ also becomes large as $T$ increases. 
This means that the $\theta$-dependence of $M_{\pi}$ 
becomes stronger as $T$ increases. 
It is then preferable that LQCD simulations will be made at 
higher temperature near 
$T_\mathrm{c}$ except the dip temperature, 
in order to determine the parameters of the PNJL model more accurately.
The $\theta$-dependence of meson masses at lower $T$ 
can be extracted from LQCD data at higher $T$ by using the PNJL model. 

\begin{figure}[htbp]
\begin{center}
 \includegraphics[width=0.4\textwidth]{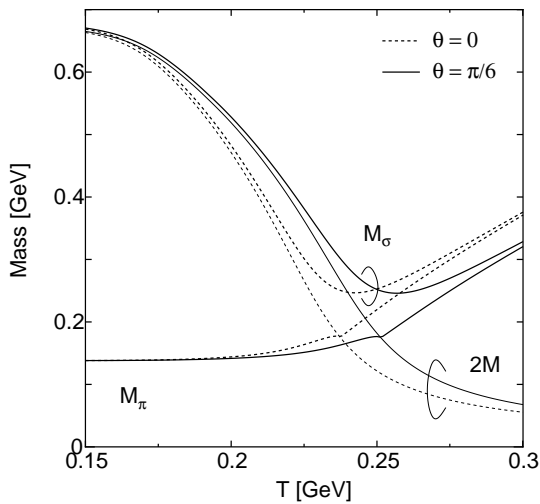}
\end{center}
\caption{$T$-dependence of sigma and pion masses
at $\theta=0$ and $\pi/6$.
Twice the constituent quark mass $M$ is also shown.}
\label{Fig-T}
\end{figure}

If $A=0$, $M_{\xi}$ will have no $\theta$ dependence, as shown by 
\eqref{fit}. The fact that 
$A/C$ is small, then, indicates 
that the $\theta$ dependence of $M_{\xi}$ is rather 
weak. Figure~\ref{Fig-T} presents the $T$-dependence of pion mass
at $\theta=0$ and $\pi/6$, where 
results of $\theta=\pi/6$ ($\theta=0$) are represented by 
solid (dotted) curves. 
As expected, the overall behavior is almost the same 
between the two cases and the difference between the two is not large, 
although the crossing point where $M_\pi=2M$ is
shifted to higher $T$ as $\theta$ increases.

The extrapolation of $M_{\xi}$ 
from $\mu^2 < 0 $ to $\mu^2 \ge 0 $ 
was  necessary for LQCD at least so far. Actually, 
LQCD data at $\mu^2 < 0 $  were extrapolated to 
the $\mu^2 \ge 0 $ region by assuming some fitting 
functions~\cite{FP,Elia2,Elia1}. 
In the PNJL model, physical quantities are calculable directly 
in the $\mu^2 \ge 0 $ region, if the parameter set is 
determined from LQCD data 
in the $\mu^2 \le 0 $ region. 
Here we assume the present PNJL result as a result of 
this $\theta$-matching approach 
and compare it with results of usual extrapolations in 
order to test the validity of the extrapolations. 
We consider two extrapolations. 
The first one is 
\begin{align}
M_{\pi}^{\mathrm{fit}} (\mu/T )
&= A(T) \cosh({3 \mu/T}) + C(T) . 
\label{fit-C}
\end{align}
When $\mu=i\mu_\mathrm{I}$, Eq.~\eqref{fit-C} equals to 
Eq.~\eqref{fit}. This form of $M_{\pi}^{\mathrm{fit}}$ 
has an exponential $\mu$-dependence at $\mu=\mu_\mathrm{R}$. 
This extrapolation is then called the exponential extrapolation 
in this paper; a similar form is used in Ref.~\cite{Elia1}
for thermodynamical quantities rather than meson masses.  
The second is the polynomial extrapolation~\cite{FP,Elia2} in which 
a polynomial of $\mu^2$ is taken up to eighth order.

In Fig.~\ref{Fig-5}, $M_{\pi}$ calculated with 
the exponential and the polynomial extrapolation 
are compared with the result of the PNJL model. 
In the two simple extrapolations, their parameters are fitted to 
the PNJL result for $\mu^2 \leq 0$ at $T=$ 160 MeV. 
At $\mu^2 \leq 0$, $M_{\pi}^{\mathrm{fit}}$ agrees with 
$M_{\pi}$ calculated with the PNJL model within thickness of curves. 
This means that the $a_k$ for $k \ge 2$ are tiny in the expansion 
\eqref{expansion}. 
The two simple extrapolations give almost the same result for 
$\mu^2 > 0 $. 
Thus, the polynomial extrapolation often used so far is essentially 
equal to the exponential one. 
The PNJL result and the results of the two extrapolations 
coincide accurately up to 
$\mu_\mathrm{R}/T \simeq$ 1.

\begin{figure}[htbp]
\begin{center}
 \includegraphics[width=0.4\textwidth]{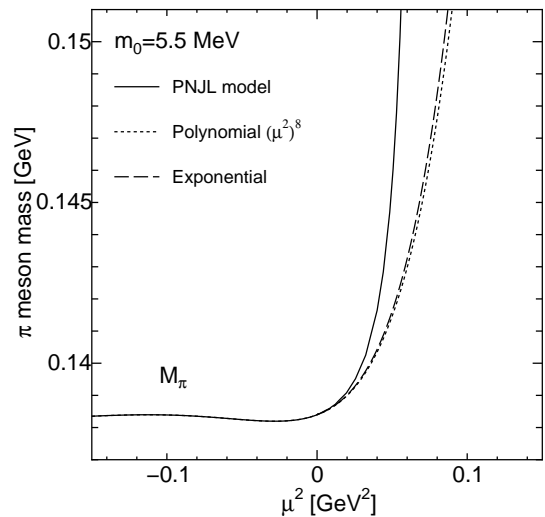} 
\end{center}
\caption{$\mu^2$-dependence of 
$M_\pi$ and $M_{\pi}^{\mathrm{fit}}$ at T=160 MeV.
The solid curve represents the PNJL result,
the dotted curve does a fitted one 
by a polynomial up to $(\mu^2)^8$ and
the dashed curve does a fitted one by \eqref{fit}.}
\label{Fig-5}
\end{figure}

Next, we discuss mathematically on 
the the analytic continuation of $M_{\xi}$ from 
$\mu=i\mu_{\rm I}$ to $\mu=\mu_{\rm R}+i\mu_{\rm I}$. 
First, we assume that LQCD gives $M_{\xi}$ numerically 
in the $\mu_{\rm I}$ region. 
In order to make the analytic continuation, 
we need an analytic form of $M_{\xi}$ that is valid 
in the $\mu_{\rm I}$ region. 
The Fourier expansion series \eqref{expansion} gives 
such an exact form of $M_{\xi}$, since $M_{\xi}$ 
is a $\theta$-even function with the RW periodicity. 
As shown below, $M_{\xi}$ has no $\theta$ dependence 
in the low-$T$ limit and then all the $a_k$ except $a_0$ tend to 
zero in the limit. 
The partition function $Z_{\rm GC}(\theta)$ with a finite value of $\theta$ is 
equivalent to $Z_{\rm GC}(0)$ with the boundary condition 
$q(x,1/T)=-\exp{(i\theta)}q(x,0)$ for the quark field $q$~\cite{RW}. 
In the low-$T$ limit where a period $1/T$ of the imaginary time becomes 
infinite, the value of $Z_{\rm GC}(0)$ does not depend on how to take 
the boundary condition 
and then has no $\theta$ dependence. 
The PNJL model can reproduce this property, as proven in Ref.~\cite{Sakai}.

As a result of the analytic continuation from 
$\mu=i\mu_{\rm I}$ to $\mu=\mu_{\rm R}+i\mu_{\rm I}$, 
the series becomes 
\bea
M_{\xi}(\mu)=\sum_{k=0} a_k(T) \cosh{(3k\mu/T)} .
\label{expansion-R} 
\eea 
This is a natural extension of the exponential extrapolation 
\eqref{fit-C}. 
This form is valid, if the series converges. 
The condition for the convergence is 
\bea
R \equiv \lim_{k \to \infty} \Big|{a_{k+1}(T) \cosh{(3(k+1)\mu/T)} 
\over a_k(T) \cosh{(3k\mu/T)} }\Big| 
< 1 .
\label{convergence} 
\eea 
In general, $R$ depends on $\mu$ and $T$. 
Now we consider the case of $\mu=\mu_{\rm R}$. 
In the low-$T$ limit, all the $a_k$ except $a_0$ tend to zero, 
as mentioned above, while the factor 
$\cosh{(3(k+1)\mu_{\rm R}/T)}/\cosh{(3k\mu_{\rm R}/T)}$ diverges 
for all $k$. 
Thus, there is a possibility that $R$ is less than 1 even in the 
low-$T$ limit. However, it is impossible to evaluate the 
extremely small coefficients 
$a_k$ ($k \ge 1$) from LQCD data in the finite $\theta$ region. 
Also for large $T$ near $T_{\rm c}$, the PNJL calculation shows that 
the coefficients $a_k$ with $k \ge 2$ are negligibly small. 
It is then difficult to evaluate the small coefficients 
from LQCD data with finite errors in the finite $\theta$ region. 
Hence we propose to use the $\theta$-matching approach 
based on the PNJL model 
instead of the exponential and polynomial extrapolations 
and the analytic continuation 
\eqref{expansion-R}.

Finally, we check an influence of the current quark mass $m_0$ on 
the $\mu$-dependence of $M_{\pi}$, since 
large $m_0$ is taken in LQCD simulations at 
finite $\theta$. 
The result adopting $m_0=$ 80 MeV is shown in Fig.~\ref{Fig-5}, for example. 
This figure indicates that, with heavy $m_0$, the sign of $dM_\pi/d\mu^2$ 
at $\mu^2=0$ becomes opposite to the case of light $m_0$ shown in 
Fig. \ref{Fig-1}. This should be noticed.

\begin{figure}[htbp]
\begin{center}
 \includegraphics[width=0.4\textwidth]{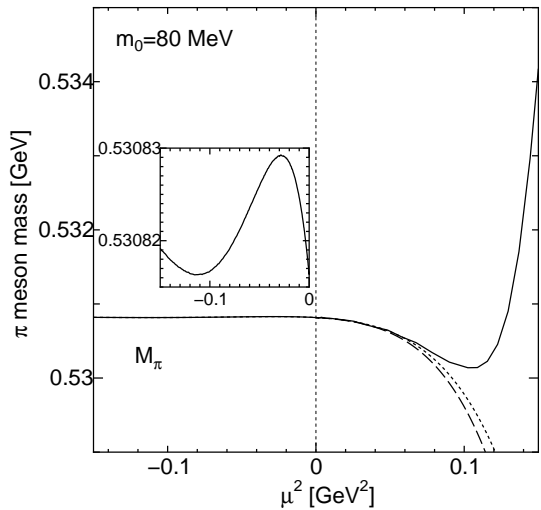} 
\end{center}
\caption{ $\mu^2$-dependence of pion mass at $T=160$ MeV in the case of 
$m_0=80$~MeV.
Definition of curves is the same as in Fig.~\ref{Fig-3}. 
}
\label{Fig-5}
\end{figure}

\section{Summary}

We have analyzed, by using the PNJL model, 
the $\mu$-dependence of pi and sigma meson masses 
in both the $\mu_\mathrm{R}$ and $\mu_\mathrm{I}$ regions. 
In the $\mu_\mathrm{I}$ region,
the meson masses $M_{\xi}(\theta)$ ($\xi=\pi$ and $\sigma$) are even functions 
of $\theta$ with the RW periodicity: 
$M_{\xi}(\theta)=M_{\xi}(-\theta)=M_{\xi}(\theta+2k\pi/3)$. 
This property is the same as that of the chiral condensate. 
The RW periodicity indicates that in general 
$M_{\xi}(\theta)$ are oscillating with $\theta$. 
The amplitude of the oscillation becomes large as $T$ increases. 
We then recommend that LQCD calculations be done at higher $T$ 
near the pseudo-critical 
temperature $T_{c}$ of the deconfinement phase transition 
at $\mu=0$ , in order to determine the parameters of the PNJL model 
more accurately.  
The $\theta$-dependence of meson masses at lower $T$ 
can be extracted from the LQCD data at higher $T$ by using the PNJL model. 
As for pion mass $M_{\pi}(\theta)$, it should be noticed that 
the phase of the oscillation 
is rather sensitive to the value of the current quark mass $m_0$.

It is possible to do the PNJL calculation with a parameter set common between 
the $\mu_\mathrm{R}$ and $\mu_\mathrm{I}$ regions. 
Hence, we can argue that meson masses and other physical quantities in 
the $\mu_\mathrm{R}$ region can be extracted from LQCD data 
on meson mass at (1) finite $T$ and zero $\mu$ and 
(2) finite $T$ and nonzero $\mu_{\rm I}$, 
by using the PNJL calculation the parameters of which 
are fixed to the LQCD data. 
The present PNJL model has three parameters, 
$m_0$, $\Lambda$, $G_{\rm s}$, in the quark sector and 
one parameter $T_0$ in the gauge sector. 
As an extension of this minimal PNJL model, 
the vector-type four-quark interaction and/or 
the scalar-type eight-quark one are occasionally 
added to the present Lagrangian~\cite{Sakai,Sakai2}.
All the parameters can be determined from LQCD data 
at the two cases; a trial is shown in Ref.~\cite{Sakai2}. 
The polynomial and exponential extrapolations used so far 
are accurate at $\mu_{\rm R}/T \lessa 1$.
The $\theta$-matching approach based on the PNJL model 
is expected to be one of the 
most reliable methods to get physical quantities 
at finite $\mu_\mathrm{R}$, if LQCD data on $M_{\xi}(\theta)$ become
available in the near future. 
We recommend that LQCD simulations 
be done systematically with the same lattice size 
between two cases of (1) finite $T$ and zero $\mu$ 
and (2) finite $T$ and nonzero $\mu_\mathrm{I}$.

\noindent
\begin{acknowledgments}
H.K. thanks M. Imachi, H. Yoneyama and M. Tachibana for useful discussions. 
K.K. thanks T. Matsumoto for careful checks of numerical calculations.
This work has been supported in part by the Grants-in-Aid for 
Scientific Research (18540280) of Education, Science, Sports, 
and Culture of Japan. 

\end{acknowledgments}


\end{document}